\documentclass[sigconf]{acmart}
\usepackage{graphicx}  %Required

\usepackage{natbib}
\usepackage[utf8]{inputenc} % allow utf-8 input
\usepackage[T1]{fontenc}    % use 8-bit T1 fonts
\usepackage{hyperref}       % hyperlinks
\usepackage{url}            % simple URL typesetting
\usepackage{booktabs}       % professional-quality tables
\usepackage{amsfonts}       % blackboard math symbols
\usepackage{nicefrac}       % compact symbols for 1/2, etc.
\usepackage{microtype}      % microtypography
\usepackage{xcolor}         % colors
\usepackage{color}
\usepackage{bm}
\usepackage{subfigure}
\usepackage{amssymb}
\usepackage{amsmath}
\usepackage{amsthm}
\usepackage[ruled,vlined]{algorithm2e}
\usepackage{array}
\usepackage{tabularx}
\usepackage{multirow}
\usepackage{verbatim}

\newcommand{\partitle}[1]{\vspace{2mm}\noindent \textbf{#1.}}
\usepackage[Symbol]{upgreek}
\usepackage{stfloats}
\usepackage{diagbox}

\setcopyright{rightsretained}
\begin{document}

% \copyrightyear{2020} 
% \acmYear{2020} 
% \setcopyright{acmcopyright}
% \acmConference[SIGIR '20] {43rd International ACM SIGIR Conference on Research and Development in Information Retrieval}{July 25--30, 2020}{Virtual Event, China}
% \acmBooktitle{43rd International ACM SIGIR Conference on Research and Development in Information Retrieval (SIGIR '20), July 25--30, 2020, Virtual Event, China}
% \acmPrice{15.00}
% \acmDOI{10.1145/XXXXXX.XXXXXX}
% \acmISBN{978-1-4503-8016-4/20/07} 
% \fancyhead{}
\setcopyright{acmcopyright}
\copyrightyear{2018}
\acmYear{2018}
\acmDOI{XXXXXXX.XXXXXXX}
\acmConference[Conference acronym 'XX]{Make sure to enter the correct conference title from your rights confirmation email}{June 03--05, 2018}{Woodstock, NY}
\acmPrice{15.00}
\acmISBN{978-1-4503-XXXX-X/18/06}
\fancyhead{}

\title{IFA: Interaction Fidelity Attention for Entire Lifelong Behaviour Sequence Modeling}

\author{Wenhui Yu}
\affiliation{%
  \institution{Kuaishou Technology}
  \city{Beijing}
  \state{China}
}
\email{yuwenhui07@kuaishou.com}

\author{Chao Feng}
\affiliation{%
  \institution{Kuaishou Technology}
  \city{Beijing}
  \state{China}
}
\email{fengchao08@kuaishou.com}

\author{Yanze Zhang}
\affiliation{%
  \institution{Kuaishou Technology}
  \city{Beijing}
  \state{China}
}
\email{zhangyanze@kuaishou.com}

\author{Lantao Hu}
\affiliation{%
  \institution{Kuaishou Technology}
  \city{Beijing}
  \state{China}
}
\email{hulantao@kuaishou.com}

\author{Peng Jiang}
\affiliation{%
  \institution{Kuaishou Technology}
  \city{Beijing}
  \state{China}
}
\email{jiangpeng@kuaishou.com}

\author{Han Li}
\affiliation{%
  \institution{Kuaishou Technology}
  \city{Beijing}
  \state{China}
}
\email{lihan08@kuaishou.com}

\iffalse
\author{Kun Gai}
\affiliation{%
  \institution{Kuaishou Technology}
  \city{Beijing}
  \state{China}
}
\email{gai.kun@qq.com}
\fi

\begin{abstract}
The lifelong user behavior sequence provides abundant information of user preference and gains impressive improvement in the recommendation task, however increases computational consumption significantly. To meet the severe latency requirement in online service, a short sub-sequence is sampled based on similarity to the target item. Unfortunately, items not in the sub-sequence are abandoned, leading to serious information loss. 

In this paper, we propose a new efficient paradigm to model the full lifelong sequence, which is named as \textbf{I}nteraction \textbf{F}idelity \textbf{A}ttention (\textbf{IFA}). In IFA, we input all target items in the candidate set into the model at once, and leverage linear transformer to reduce the time complexity of the cross attention between the candidate set and the sequence without any interaction information loss. We also additionally model the relationship of all target items for optimal set generation, and design loss function for better consistency of training and inference. We demonstrate the effectiveness and efficiency of our model by off-line and online experiments in the recommender system of Kuaishou. 

\end{abstract}



% \keywords{Negative sampling, label noise robust learning, item recommendation.}

%\settopmatter{printacmref=false}
\maketitle

\section{Introduction}
Recommender system learns preference based on history records and predicts preferred items for each user. It is important in all online platforms due to the efficient information matching. In most recomenders, users and items are represented by dense embeddings and the match scores are derived by inner product \cite{MF,BPR,FM} or fully-connected layers \cite{NCF,NFM,DeepFM}. To learn better preference, more information \cite{FM,VBPR,AES,text1,text2,GNN_Social,knowledge_base} and complex structures \cite{light_GCN,GNN_WebScale,LCFN,LGCN,nbpo,he_Attentive,sasrec,recgpt} are used. 

In industry recommendation, user behaviour sequence is very important \cite{din,dien,sim,twin}. Deep Interest Network (DIN) learns interactions between target item and behaviour sequence items by target attention \cite{din,dien}. In real-world application, we can improve the accuracy of the model significantly by increasing the length of the sequence, yet the computational consumption also increases. To reduce the complexity, Search-based Interest Model (SIM) separates target attention into two-step concatenation \cite{sim}: (1) A simple General Search Unit (GSU) is used to select a sub-sequence from the long sequence base on certain similarity rule to the target item, such as the same category (hard search) or embedding similarity (soft search). (2) An Exact Search Unit (ESU) is then used for target attention. In this paradigm, interactions between the target item and the sequence items which are not in the sub-sequence are missed, nevertheless these interactions are also very important. We give an example to demonstrate this point.

\begin{figure*}[ht]
	\centering
	\subfigure[Target item]{
		\includegraphics[scale = 0.34]{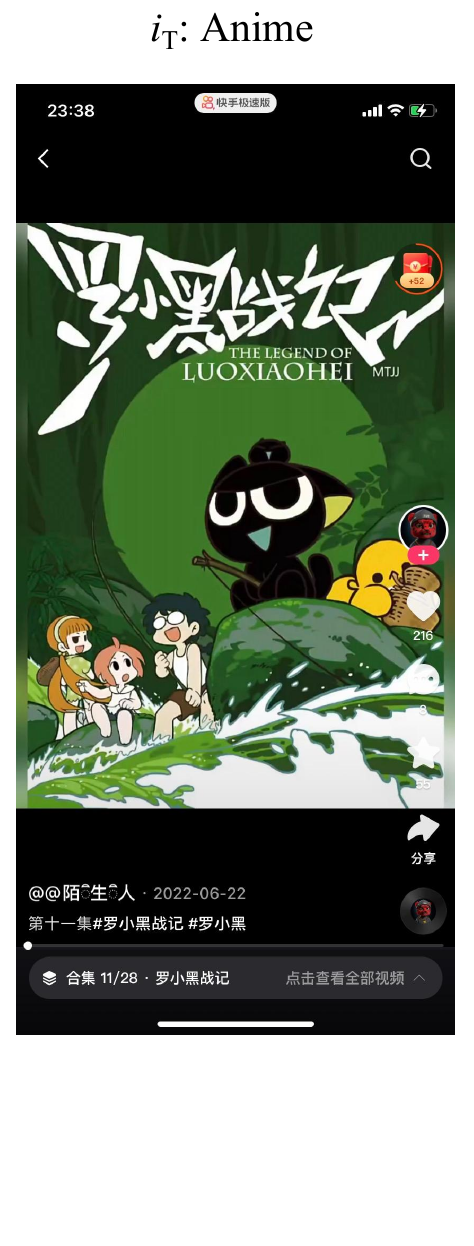}
		\label{subfig:target_item}
	}
	\hspace{1mm}
	\subfigure[Long sequence]{
		\includegraphics[scale = 0.34]{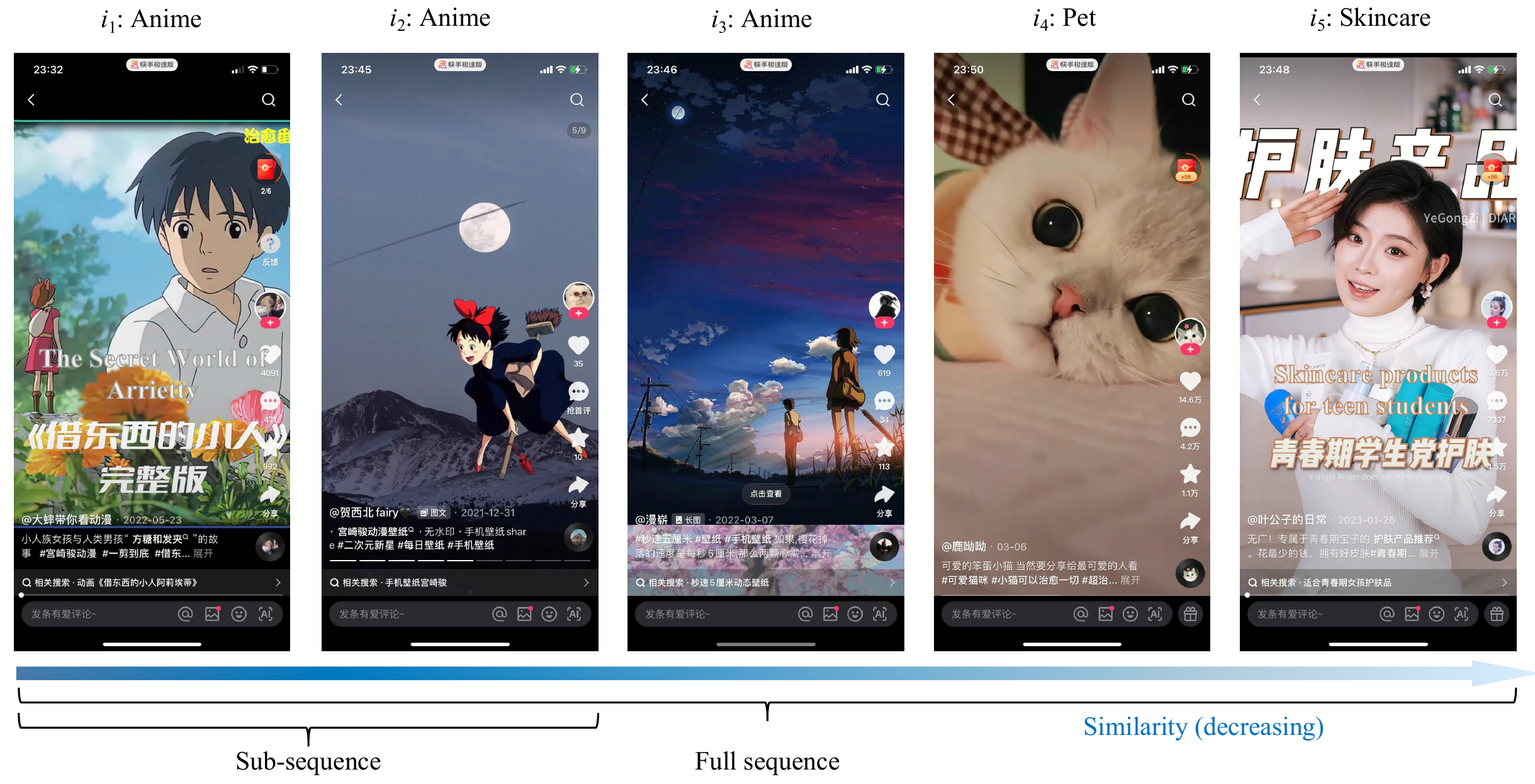}
		\label{subfig:long_seq}
	}
	\caption{Sampling mechanism in SIM}
	\label{fig:sample_sub_seq}
\end{figure*}

Figure \ref{fig:sample_sub_seq} shows the sequence sampling mechanism in SIM. Figure \ref{subfig:target_item} is the target time $i_T$, which is an anime. Figure \ref{subfig:long_seq} is the user lifelong sequence $\{i_1, i_2, i_3, i_4, i_5\}$, including anime, pet, and skincare categories. Taking hard search as example, assuming the sub-sequence is $\{i_1, i_2\}$,  interactions of $i_T$ and $\{i_3, i_4, i_5\}$ are missed in this case. Unfortunately, these interactions are important, and we will illustrate in tree aspects:

\begin{itemize}
{\item The similarity rule of GSU is simple, hence sequence items which are relevant to the target item may be missed. For example, $i_4$ has the same topic (cat) with $i_T$, yet is discarded due to the different categories. Though complex GSUs are proposed to alleviate this issue, such as soft search \cite{sim} or a lightweight target attention \cite{twin}, they cannot solve it completely since GSU has to trade off the accuracy and efficiency.}

{\item Sequence items which are irrelevant to the target item may also provide important information of user profile. For example, from $i_5$ we can infer that the user may be a young woman, with good education and aesthetic taste.}

{\item Restricted by length of the sub-sequence, sequence items which are similar to the target item may also be discarded. For example, assuming the max length of the sub-sequence is 2, we  have to discard $i_3$ even it is also in the anime category.}
\end{itemize}

We come to the conclusion that the sampling paradigm damages abundant interaction information and hurt the performance of the model seriously. For better interaction fidelity, we propose a new paradigm, which reserves the whole sequence. To design a new paradigm, we first analyze why the existing paradigm has to damage interaction information to simplify computation. We argue that the reason is that only one target item is predicted for each time, thus the complexity of attention with a $n$-length sequence is $O(n)$. The optimization space of linear complexity is small, thus the simplification strategy is very limited, and causes information loss inevitably. A $k$-length sub-sequence is sampled to reduce the complexity to $O(k)\sim O(\log n)$ in SIM, and interaction information is lost.

However, there are $m$ target items in the candidate set to be scored in the serving stage, hence we need to perform the $O(n)$ computation $m$ times. Inspired by this, we input the whole candidate set into the model for one time to achieve quadratic time complexity $O(mn)$ without any extra computation. Now we have larger optimization space and more choices. To perform cross attention of the candidate set and behaviour sequence efficiently, we adopt linear transformer \cite{linearTransformers} to reduce the complexity to $O(m+n)$ without any interaction information loss. In this case, inferring one target item costs $O(1+n/m)$ time, which is a constant if $m$ and $n$ are similar. We call our model \textbf{I}nteraction \textbf{F}idelity \textbf{A}ttention (\textbf{IFA}), which is a new paradigm for full long behaviour sequence modeling.

As we need to input the whole candidate set into IFA, thus we additionally propose two optimizations: (1) Ranking task is to generate an optimal $m'$-size candidate set from the $m$-size one, where $m' \ll m$. Considering that widely-used ranking models \cite{din,dien,sim,twin} learn a set of $m'$ top-1 items rather than a best top-$m'$ set in a point wise way, we design an efficient relationship aware module to model the relationship for large-scale candidate sets. (2) We solve the Sample Selection Bias (SSB) issue \cite{ssb} by training and inferring on candidate set. To learn better probability contribution, we model \textit{cli}ck through \textit{imp}ression rate $pCTR=p(cli|imp)$ and \textit{imp}ression through \textit{can}didate rate $pITR=p(imp|can)$, and combine them into impression through \& click through rate $pITCTR=p(cli|can)$.

The contributions are listed as follows:

\begin{itemize}
{\item We propose an Interaction Fidelity Attention (IFA) model to construct interactions between target items and sequence items without any information loss efficiently.}

{\item We model the relationship between target items in the candidate set to return an optimal top-$m'$ set.}

{\item We design a loss function for better consistency of training and inference.}

{\item We design comprehensive experiments to demonstrate the effectiveness of our proposed IFA model.}
\end{itemize}

\section{Related Work}
In this section, we introduce some related work about lifelong sequence ranking models, efficient transformer, and relationship aware models.
\subsection{Lifelong Sequence Modeling}
Behaviour sequence provides abundant information of user interests and is widely-used in industry recommendation models. \citet{din,dien} perform attention with the target item as query and short-term user behavior sequence as key and value, to search if there are historical items similar to the target item. Real-world applications show that the performance increases significantly with the increasing of sequence length. To involve in lifelong sequence, SIM and variants are proposed by separating the target attention into a concatenation of GSU (to sample sub-sequence) and ESU (to build interactions of target item and sub-sequence) \cite{sim,twin}. GSU is a simple similarity rule, such as in the same category or cluster \cite{sim} and a lightweight target attention \cite{twin}. All existing models for lifelong sequence adopt this two-step structure thus lose interaction information. In this paper, we design a new paradigm to reserve all items in the sequence.

\subsection{Efficient Transformer}
Our IFA model needs efficient transformer to simplify quadratic-complexity attention. There are three paths: (1) sampling interactions \cite{reformer,longformer,bigbird}; (2) building coarse-grained interactions by clustering centers \cite{clusformer1,clusformer2,setformer,Linformer}; and (3) using kernel function to replace softmax and changing the order of matrix multiplication \cite{linearTransformers}. Considering paths 1 and 2 both damage the interaction information (coverage and accuracy of interaction), we adopt path 3 in IFA model. To be specific, we leverage Linear Transformer \cite{linearTransformers} to replace the standard attention.

\subsection{Relationship Aware Modeling}
We aim to generate the best top-$m'$ set in the ranking stage, however, existing models only consider if the item fit the user's preference, which generate a set of $m'$ top-1 items. To close this gap, \cite{prm} leverages transformer to model the mutual influence of target items in the candidate set. However, this method is for the re-ranking task, where the candidate set is small. In this paper, we design an efficient module for relation aware in large-scale candidate set.

\section{Preliminary}
\label{sec:preliminary}
% In this section, we introduce the preliminary about linear transformer. In this paper, we use Italic letters to indicate numbers, and use bold uppercase/lowercase letters refer to matrices/vectors. For example, ${\bm{{\rm A}}}$ is a matrix, ${\bm{{\rm A}}}^{(x)}$ is a matrix for attribute $x$, and ${\bm{{\rm A}}}_{ij}$ is the value at the $i$-th row and the $j$-th column of ${\bm{{\rm A}}}$. For input matrices of query ${\bm{{\rm E}}}^{(Q)}\in \mathbb{R}^{m \times d_Q}$, key ${\bm{{\rm E}}}^{(K)}\in \mathbb{R}^{n \times d_K}$, and value ${\bm{{\rm E}}}^{(V)}\in \mathbb{R}^{n \times d_V}$, the cross attention is defined as ${\bm{{\rm E}}}={\rm softmax}(\frac{{\bm{{\rm Q}}}{\bm{{\rm K}}}^\mathsf{T}}{\sqrt{D}}){\bm{{\rm V}}}$, where ${\bm{{\rm Q}}}={\bm{{\rm E}}}^{(Q)}{\bm{{\rm W}}}^{(Q)}$, ${\bm{{\rm K}}}={\bm{{\rm E}}}^{(K)}{\bm{{\rm W}}}^{(K)}$, and ${\bm{{\rm V}}}={\bm{{\rm E}}}^{(V)}{\bm{{\rm W}}}^{(V)}$. To reduce the time complexity, \cite{linearTransformers} proposed linear transformer by kernel method and associative property of matrix multiplication:
In this section, we introduce the preliminary about Linear Transformer. In this paper, we use Italic letters to indicate numbers, and use bold uppercase/lowercase letters refer to matrices/vectors. For example, ${\bm{{\rm A}}}$ is a matrix, ${\bm{{\rm A}}}_{x}$ is a matrix for attribute $x$. For input matrices of query ${\bm{{\rm E}}}_Q\in \mathbb{R}^{m \times d_Q}$, key ${\bm{{\rm E}}}_K\in \mathbb{R}^{n \times d_K}$, and value ${\bm{{\rm E}}}_V \in \mathbb{R}^{n \times d_V}$, the cross attention is defined as ${\bm{{\rm E}}}={\rm softmax}(\frac{{\bm{{\rm Q}}}{\bm{{\rm K}}}^\mathsf{T}}{\sqrt{d}}){\bm{{\rm V}}}$, where ${\bm{{\rm Q}}}={\bm{{\rm E}}}_Q{\bm{{\rm W}}}_Q$, ${\bm{{\rm K}}}={\bm{{\rm E}}}_K{\bm{{\rm W}}}_K$, and ${\bm{{\rm V}}}={\bm{{\rm E}}}_V{\bm{{\rm W}}}_V$. To reduce the time complexity, \citet{linearTransformers} proposed Linear Transformer by kernel method and associative property of matrix multiplication:
\begin{flalign}
\label{equ:linear_transformer}
{\bm{{\rm E}}}=\left(\phi({\bm{{\rm Q}}})\phi({\bm{{\rm K}}})^\mathsf{T}\right){\bm{{\rm V}}} = \phi({\bm{{\rm Q}}})\left(\phi({\bm{{\rm K}}})^\mathsf{T}{\bm{{\rm V}}}\right),
\end{flalign}
where $\phi(\cdot\;)$ is the row-wise kernel function to approximate softmax.

\section{Interaction Fidelity Attention Model}
In this section, we introduce the IFA model for lifelong sequence.

\subsection{Problem Formulation}
For certain user $u$, we have a candidate set $\mathcal{C}_u = \{i_1, i_2, \cdots, i_m\}$ returned from the retrieve stage, and a long user behaviour sequence $\mathcal{S}_u = \{j_1, j_2, \cdots, j_n\}$. In point-wise model such as DIN, we aim to give a score $y_{i}$ for each target item $i \in \mathcal{C}_u$ based on the user-wide feature $f_u$ (such as user id, age, gender, etc.), item-wide feature $f_i$ (such as item id, category, author id, etc.), cross feature $f_{ui}$, and behaviour sequence $\mathcal{S}_u$: $y_{i}={\rm DIN}(f_u, f_i, f_{ui}, \{f_j\}_{j\in \mathcal{S}_u})$. For long sequence, SIM uses a GSU to cut off $\mathcal{S}_u$ based on the similarity to $i$: $y_{i}={\rm SIM}(f_u, f_i, f_{ui}, \{f_j\}_{j\in \mathcal{S}_u})={\rm DIN}(f_u, f_i, f_{ui}, \{f_j\}_{j\in {\rm GSU}(\mathcal{S}_u, i)})$. In our IFA, we input the whole candidate set into the model: $\{y_{i}\}_{i\in \mathcal{C}_u}={\rm IFA}(f_u, \{f_i, f_{ui}\}_{i\in \mathcal{C}_u}, \{f_j\}_{j\in \mathcal{S}_u})$, and simplify cross attention to reduce the complexity.

\subsection{Model Structure}

\partitle{Embedding Layer} We encode the categorical features $f_x$ of attribute $x$ into trainable embedding vectors ${\bm{{\rm f}}}_x \in \mathbb{R}^{1 \times d_x}$. The item-wide feature embeddings of the whole candidate set are denoted as ${\bm{{\rm F}}}_{can}=[{\bm{{\rm f}}}_{i_1}^\mathsf{T}, {\bm{{\rm f}}}_{i_2}^\mathsf{T}, \cdots, {\bm{{\rm f}}}_{i_m}^\mathsf{T}]^\mathsf{T}\in \mathbb{R}^{m \times d_i}$. The cross feature embeddings are denoted as ${\bm{{\rm F}}}_{cro}=[{\bm{{\rm f}}}_{ui_1}^\mathsf{T}, {\bm{{\rm f}}}_{ui_2}^\mathsf{T}, \cdots, {\bm{{\rm f}}}_{ui_m}^\mathsf{T}]^\mathsf{T}\in \mathbb{R}^{m \times d_{ui}}$. The sequence embeddings are denoted as ${\bm{{\rm F}}}_{seq}=[{\bm{{\rm f}}}_{j_1}^\mathsf{T}, {\bm{{\rm f}}}_{j_2}^\mathsf{T}, \cdots, {\bm{{\rm f}}}_{j_n}^\mathsf{T}]^\mathsf{T}\in \mathbb{R}^{n \times d_i}$. And the user-wide feature embeddings are ${\bm{{\rm f}}}_u$ repeated $m$ times: ${\bm{{\rm F}}}_u=[{\bm{{\rm f}}}_u^\mathsf{T}, \cdots, {\bm{{\rm f}}}_u^\mathsf{T}]^\mathsf{T}\in \mathbb{R}^{m \times d_u}$

\partitle{Interactions Between Candidate Set and Sequence} In this part, we propose the key module of our model. We use cross attention with ${\bm{{\rm F}}}_{can}$ as query and with ${\bm{{\rm F}}}_{seq}$ as key and value to construct interactions of target items and sequence items:
\begin{flalign}
\label{equ:self_att_cs}
{\bm{{\rm F}}}_{c-s} = {\rm fastAtten}({\bm{{\rm F}}}_{can}, {\bm{{\rm F}}}_{seq}, {\bm{{\rm F}}}_{seq}),
\end{flalign}
where ${\rm fastAtten}(query, key, value)$ is the fast attention with linear complexity. In this paper, we adopt Linear Transformer \cite{linearTransformers} introduced in Section 3. 

\partitle{Interactions in Candidate Set} We also construct interactions between target items in the candidate set by self attention with ${\bm{{\rm F}}}_{can}$ as query, key, and value:
\begin{flalign}
\label{equ:self_att_cc}
{\bm{{\rm F}}}_{c-c} = {\rm fastAtten}({\bm{{\rm F}}}_{can}, {\bm{{\rm F}}}_{can}, {\bm{{\rm F}}}_{can}),
\end{flalign}
where ${\rm fastAtten}(query, key, value)$ is also Linear Transformer. 

\partitle{Fully-connected Layer} 
We concatenate all features ${\bm{{\rm F}}}=[{\bm{{\rm F}}}_{c-s}, {\bm{{\rm F}}}_{c-c}, $ ${\bm{{\rm F}}}_{u}, $ ${\bm{{\rm F}}}_{cro}] \in \mathbb{R}^{m \times d}$, and input ${\bm{{\rm F}}}$ into a multi-layer perceptron (MLP) to get scores of all candidate items: 
\begin{equation}
\label{equ:mlp}
\begin{split}
{\bm{{\rm F}}}_{l+1} &= ReLU({\bm{{\rm F}}}_l{\bm{{\rm W}}}_l + {\bm{{\rm b}}}_l), \\
{\bm{{\rm y}}} &= \sigma({\bm{{\rm F}}}_L{\bm{{\rm w}}} + b),
\end{split}
\end{equation}
where ${\bm{{\rm W}}}_l$ and ${\bm{{\rm w}}}$ are weight; ${\bm{{\rm b}}}_l$ and $b$ are bias; There are $L$ hidden layers in the MLP and ${\bm{{\rm F}}}_0={\bm{{\rm F}}}$; and ${\bm{{\rm y}}} \in \mathbb{R}^{m \times 1}$ is the prediction. To predict multiple objectives, we use several MLP networks.

\begin{figure*}[ht]
\centering
\includegraphics[scale = 0.845]{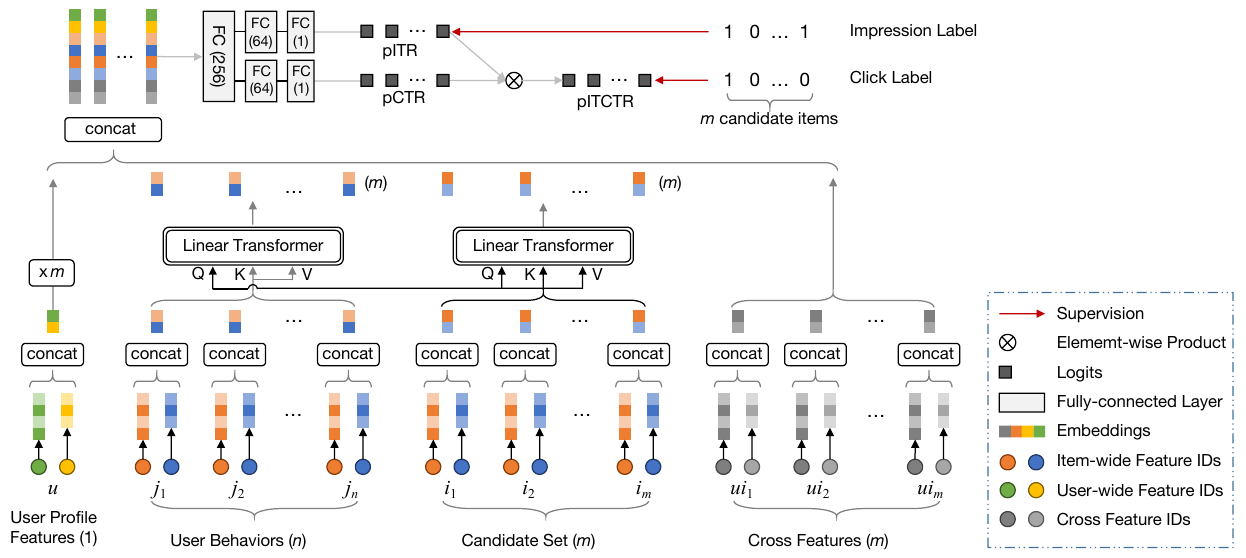}	
\caption{Illustration of IFA}
\label{fig:model_structure}
\end{figure*}

We represent our IFA in Figure \ref{fig:model_structure}. It is necessary to point out that we only introduce the most important structure for concise representation in this paper. We can add more modules in real-world application, such as self-attention of sequence, MMOE layer \cite{mmoe}, PEPnet \cite{pepnet}, LayerNorm and ResNet \cite{transformer}.

\subsection{Optimization}
\label{subsec:opt}
In this section, we introduce the optimization method. To learn the probability of all candidate items to be clicked by the user $p(cli|can)$. We follow \cite{esmm} to deal with the sparsity issue by modeling the sequential pattern of $candidate \rightarrow impression \rightarrow click$, thus we have $p(cli|can) = p(cli|imp) \times p(imp|can)$.

We use an MLP shown in Equation (\ref{equ:mlp}) as the impression tower, with the output denoted as ${\bm{{\rm y}}}_{imp}$; and use another MLP as the click tower, with the output denoted as ${\bm{{\rm y}}}_{cli}$. Impression labels are denoted as ${\bm{{\rm l}}}_{imp}\in \{0, 1\}^{m \times 1}$ and click labels are denoted as ${\bm{{\rm l}}}_{cli}\in \{0, 1\}^{m \times 1}$. The impression loss and the click loss are defined as:
\begin{equation}
\label{equ:loss}
\begin{split}
&L_{imp} = {\bm{{\rm l}}}_{imp}^\mathsf{T}\log {\bm{{\rm y}}}_{imp} + ({\bm{{\rm 1}}}-{\bm{{\rm l}}}_{imp})^\mathsf{T}\log(1-{\bm{{\rm y}}}_{imp}),\\
&L_{cli} = {\bm{{\rm l}}}_{cli}^\mathsf{T}\log ({\bm{{\rm y}}}_{imp}\odot{\bm{{\rm y}}}_{cli}) + ({\bm{{\rm 1}}}-{\bm{{\rm l}}}_{cli})^\mathsf{T}\log(1-{\bm{{\rm y}}}_{imp}\odot{\bm{{\rm y}}}_{cli}),
\end{split}
\end{equation}
where $\odot$ means element-wise product, and ${\bm{{\rm 1}}}\in \{1\}^{m \times 1}$; ${\bm{{\rm y}}}_{i}$ is to predict $p(imp|can)$; ${\bm{{\rm y}}}_{c}$ is to predict $p(cli|imp)$; ${\bm{{\rm y}}}_{i}\odot{\bm{{\rm y}}}_{c}$ is to predict $p(cli|can)$. When inference, we use $pITCTR$ to order all candidate items. 

\subsection{Fast Attention}
Fast attention in Equations (\ref{equ:self_att_cs}) and (\ref{equ:self_att_cc}) is the key component of our IFA to achieve interaction fidelity attention. There are many choices for fast transformer \cite{reformer,longformer,bigbird,Linformer,setformer,clusformer1,clusformer2,linearTransformers}. In summary, there are three methods to reduce the time complexity of transformer: sampling \cite{reformer,longformer,bigbird}, clustering \cite{Linformer,setformer,clusformer1,clusformer2} and associative property \cite{linearTransformers}. In this subsection, we discuss the effectiveness of these three methods and give our choice.

\partitle{Sampling} 
Cross attention is to construct $mn$ interactions between query (candidate set with $m$ items) and key (behaviour sequence with $n$ items), which is computationally expensive, thus an intuitive way is to sample a small part of these interactions. For example, Reformer finds similar key embeddings for each query embedding, and only builds interactions between them \cite{reformer}. Sampling methods are sub-optimal for us since our IFA degenerates into SIM.

\partitle{Clustering} 
Another way for simplification is to construct coarse-grained interactions by clustering center. One thing need to be pointed out is that besides the methods which cluster embeddings implicitly by clustering algorithm \cite{clusformer1,clusformer2}, we also regard Set Transformer \cite{setformer} and Linformer \cite{Linformer} as clustering methods since they also achieve clustering by attention and matrix multiplication. Clustering methods are also sub-optimal since IFA degenerates into MIND \cite{mind} or ComiRec \cite{comirec}.

\partitle{Associative Property} 
Sampling and clustering methods all damage the interaction information fidelity (coverage and accuracy) to reduce the time consumption. To increase interaction information fidelity, we try Linear Transformer \cite{linearTransformers}, which removes the softmax function and utilizes associative property to reduce the complexity (please see Section \ref{sec:preliminary} for details). There is no interaction information loss: each query embedding interacts with each key embedding directly and accurately, and all interactions are reserved. We take Linear Transformer as the fast attention in IFA.

However, softmax is used for normalization. Removing it directly makes aggregation explodes and training unstable. Experiments show that IFA does not converge. To address this issue, we design normalization mechanism without locking the multiplication order. Inspired by Graph Convolutional Network (GCN) \cite{yu2021self}, we use the degree matrix for normalization and rewrite Equation (\ref{equ:linear_transformer}):
\begin{flalign}
\label{equ:norm_linear_transformer}
{\bm{{\rm E}}}={\bm{{\rm D}}}^{-1}\phi({\bm{{\rm Q}}})\left(\phi({\bm{{\rm K}}})^\mathsf{T}{\bm{{\rm V}}}\right),
\end{flalign}
where ${\bm{{\rm D}}}$ is a diagonal degree matrix, and the $i$-th diagonal element is sum of the $i$-th row of the matrix $\phi({\bm{{\rm Q}}})\phi({\bm{{\rm K}}})^\mathsf{T}$. Since $\phi({\bm{{\rm Q}}})\phi({\bm{{\rm K}}})^\mathsf{T}$ cannot be calculated explicitly, we give the deduction of ${\bm{{\rm D}}}$ based on the definition: 
\begin{flalign}
\label{equ:degree_matrix}
{\bm{{\rm D}}}={\rm diag}\left(\phi({\bm{{\rm Q}}})\left(\phi({\bm{{\rm K}}})^\mathsf{T}{\bm{{\rm 1}}}\right)\right),
\end{flalign} where ${\bm{{\rm 1}}} \in \{1\}^{n\times 1}$ is used to reduce sum $\phi({\bm{{\rm K}}})$ by multiplication $\phi({\bm{{\rm K}}})^\mathsf{T}{\bm{{\rm 1}}}$, and diag( ) is to generate a diagonal matrix. Please note that we need to select $\phi(\;)$ to make sure that each element of ${\bm{{\rm D}}}$ is non negative, such as softplus or ReLU.

\subsection{Time Complexity}
IFA and SIM are both simplification of DIN. We give the time complexity of processing one target item to demonstrate the efficiency of our proposed method. For DIN, the time consumption of attention with one target item as query and $n$-length sequence as key and value is $O(n)$. For SIM, the time consumption is $O(k)$, where $k\sim \log n$ is the length of the sub-sequence. 

For IFA, if we use standard attention, the time complexity of processing $m$ target items are $O(mn)$, and we reduce the time complexity to $O(m+n)$ by linear transformer. In this case, processing one target item costs $O(1+n/m)$ time. In prerank, $m\sim 10$k and $n\sim 10$k, thus the time complexity is a constant.

\section{Experiments}
In this section, we conduct experiments to demonstrate the effectiveness of our IFA model by comparing it with several state-of-art (SOTA) models on Kuaishou industry data and by online AB test.

\subsection{Experimental Setup}
In this section, we introduce experimental setups, including datasets, baselines, and evaluation protocols. 

\partitle{Datasets} Considering that we need samples with both candidate set and long sequence, we adopt Kuaishou online streaming data for experiment since no published dataset meets the requirements. As the quantity of candidate items are much larger than that of impressed items, we sample 10\% requests as the training samples.

\begin{table}[ht]
    \caption{Statistics of the streaming data on Kuaishou in a day (10\% requests)}
    \begin{center}
    \scalebox{1.0}{
        \begin{tabular}{ll}
            \toprule
            Attributes  & Quantity       \\ 
            \midrule
            User        & 144,328,769          \\
            Item        & 70,799,315      \\
            Request     & 300,713,382      \\
            Candidate   & 756,527,143,866      \\
            Impression  & 980,126,567     \\
            Click       & 608,213,881      \\
            Long view   & 478,241,795      \\
            Like        & 81,806,840        \\
            \bottomrule
        \end{tabular}}
    \end{center}
    \label{tab:datasets}
\end{table}

Besides click, we also predict other user actions such as long view and like. We train the model on the streaming data to predict impression from candidate and predict user actions from impression. For each batch of the data, we test the model first for evaluation, and then train the model.

\partitle{Baselines} We compare the following models to demonstrate the effectiveness of our IFA model.

\begin{itemize}
    \item \textbf{AvgPooling:} This method processes user behaviour sequence by \textbf{Av}era\textbf{g}e \textbf{Pooling}. It is a simple yet effective way to model short behaviour sequence.
    \item \textbf{DIN:} This \textbf{D}eep \textbf{I}nterest \textbf{N}etwork achieves user-item interaction by performing attention between the target item and the user behaviour sequence \cite{din}. DIN is the SOTA model for short behaviour sequence.
    \item \textbf{SIM-hard:} This \textbf{S}earch-based \textbf{I}nterest \textbf{M}odel with \textbf{hard} search uses a two-step interaction module instead of the end-to-end attention to reduce the complexity \cite{sim}. In the first step, GSU generates a sub-sequence by selecting sequence items with the same category to the target items, and in the second step, ESU conducts the standard attention.
    \item \textbf{SIM-soft:} In this \textbf{S}earch-based \textbf{I}nterest \textbf{M}odel with \textbf{soft} search, GSU defines similarity by the distance of embeddings \cite{sim}.
    \item \textbf{TWIN:} This \textbf{TW}o-stage \textbf{I}nterest \textbf{N}etwork uses a light attention as GSU for better consistency with ESU \cite{twin}. TWIN is the SOTA model for lifelong behaviour sequence.
\end{itemize}

In our experiments, the length of the lifelong sequence $n=10,000$ and the length of the sub-sequence/short sequence $k=100$.

\partitle{Evaluation Protocols}
For each sample in the test step, we rank all candidate items for the user and return the top items. We apply area under curve (AUC) to evaluate the recommendation quality. In the test samples, items with user actions are labeled as positive and candidate items without user actions are labeled as negative. In another words, we test the models' ability of predicting user actions from the whole candidate set. 

As that baselines are trained on impressed items traditionally yet IFA is trained on candidate items, IFA has a natural advantage in this task. For fair comparison, we train all models on the scope of candidate set, and all models in the experiment are optimized by ESMM \cite{esmm} introduced in Section \ref{subsec:opt}.

\subsection{Performance of IFA Model}
In this section, we compare our IFA models against baselines on online streaming data of Kuaishou and report the result in Table \ref{tab:all_models}.  

\begin{table}[ht]
    \caption{Performance of all models}
    \begin{center}
    \scalebox{1.0}{
        \begin{tabular}{lcccc}
            \toprule
            Model       & Impression &  Click & Long view & Like \\ 
            \midrule
            AvgPooling  & 0.9361 & 0.9444 & 0.9523 & 0.9506 \\
            DIN         & 0.9409 & 0.9498 & 0.9564 & 0.9561 \\
            SIM\_hard   & 0.9452 & 0.9556 & 0.9605 & 0.9651 \\
            SIM\_soft   & 0.9502 & 0.9566 & 0.9642 & 0.9657 \\
            TWIN        & 0.9582 & 0.9632 & 0.9726 & 0.9737 \\
            \midrule
            IFA         & \textbf{0.9788} & \textbf{0.9815} & \textbf{0.9864} & \textbf{0.9808} \\
            \bottomrule
        \end{tabular}}
    \end{center}
    \label{tab:all_models}
\end{table}

As shown in the table, AvgPooling achieves shallow user-item interaction hence performs the worst in all models. By performing attention between the target item and behaviour sequence, DIN achieves deep user-item interaction and performs better than AvgPooling. Benefiting from long-term user preference, SIM outperforms DIN obviously. SIM\_soft and SIM\_hard show similar performance and SIM\_soft performs better in most cases. TWIN improves the consistency of GSU and ESU, and achieves further performance improvement.

Though some baselines show competitive performance in experiment, all baselines lose information of the lifelong sequence. By modeling the sequence without any interaction information loss, the proposed IFA achieves the best performance significantly. IFA outperforms the most competitive baseline TWIN 2.06pp for the best case.

\subsection{Ablation Study}
In this section, we demonstrate the effectiveness of each component in IFA. There are two important components of IFA:

\begin{itemize}
    \item Full Sequence Modeling (FSM) mechanism shown in Equation (\ref{equ:self_att_cs}). FSM is the key component of the IFA model. We expand the target item to a candidate set, and then perform efficient cross attention to achieve lossless simplification. To demonstrate the effectiveness of FSM, we replace it with the two-step attention proposed in \cite{twin} in the ablation experiment.
    \item Relation Aware Modeling (RAM) mechanism shown in Equation (\ref{equ:self_att_cc}). RAM is proposed to model the relationship of the candidate items to generate the best recommendation set. To demonstrate the effectiveness of RAM, we remove it in the ablation experiment.
\end{itemize}

\begin{table}[ht]
    \caption{Ablation study of IFA}
    \begin{center}
    \scalebox{1.0}{
        \begin{tabular}{lcccc}
            \toprule
            Model       & Impression &  Click & Long view & Like \\ 
            \midrule
            IFA         & \textbf{0.9788} & \textbf{0.9815} & \textbf{0.9864} & \textbf{0.9808} \\
            IFA-FSM     & 0.9652 & 0.9691 & 0.9774 & 0.9752  \\
            IFA-FSM-RAM & 0.9582 & 0.9632 & 0.9726 & 0.9737  \\ 
            \bottomrule
        \end{tabular}}
    \end{center}
    \label{tab:ablation}
\end{table}

The result of ablation study is shown in Table \ref{tab:ablation}. As we can see, without FSM, the performance decreases significantly since we fail to model the whole lifelong sequence. By further removing RAM, the performance also decreases since by predicting the preference on each item in a point wise way, we generate the recommendation set greedily, and the result is a local optimal solution. Please note that IFA-FSM-RAM in Table \ref{tab:ablation} is TWIN in Table \ref{tab:all_models}.

\subsection{Performance of AB test}
We apply IFA by adding a new model in the industrial recommendation service of Kuaishou, which serves billions of users. We conduct an AB test and report the relative improvements in Table \ref{tab:ab_test}. To show the significance, we also provide acceptance region with p-value is 0.05, and confidence interval is improvement + acceptance region. As represented in the table, IFA achieves significant improvement in many important metrics, especially watch time. IFA not only improves the accuracy, but also improves the diversity significantly. The reason is that compared with SIM, interaction in IFA are not constrained by the category, thus target item with a new category can be returned.

\begin{table}[ht]
\caption{Performance of AB test}
\label{tab:ab_test}
\centering
\scalebox{1.0}{
\begin{tabular}{p{1.7cm}p{2.0cm}p{3.0cm}}
    \toprule
    Objective    & Improvement & Acceptance region \\
    \midrule
    Watch time & +0.803\% & [-0.10\%, 0.10\%]  \\
    Impression & +0.291\% & [-0.14\%, 0.14\%]  \\
    Like       & +0.984\% & [-0.36\%, 0.37\%]  \\
    Comment    & +1.329\% & [-0.56\%, 0.56\%]  \\
    Follow     & +1.781\% & [-0.49\%, 0.49\%]  \\
    Share      & +0.780\% & [-0.86\%, 0.86\%]  \\
    Diversity  & +0.821\% & [-0.05\%, 0.05\%]  \\
    \bottomrule
\end{tabular}}
\end{table}

Based on these promising results, we fully launched our IFA. We launched three versions by increasing $n$ continuously, and report the accumulate improvement in Table \ref{tab:ab_test}. The AB test is observed for 7 days (2023.11.24-2023.11.29) on 10\% users or 3 days (2024.02.03-2024.02.05 and 2024.06.10-2024.06.12) on 20\% users to get reliable result.

\section{Conclusion and Future Work}
In this paper, we design an efficient way to model the whole lifelong behaviour sequence in the recommendation task called IFA. To achieve this, we expand the query of attention from a single embedding into an embedding set, and leverage linear transformer to achieve efficient cross attention without any interaction information loss. We also additionally construct the relationship between items in the candidate set to generate the recommendation set as an optimal whole. Offline experiments and online AB test show the effectiveness of IFA.

For future work, we are interested in solving longer sequence or larger candidate set in the recommendation task. We also have interests in  modeling the sequential information of the behaviour sequence by GPT paradigm in IFA. Exploring the power of IFA in more tasks such as search is also a way worth trying .

\clearpage

\newpage
\bibliographystyle{ACM-Reference-Format}
\bibliography{_sample-bibliography}

\end{document}